\input phyzzx
\overfullrule=0pt
\def\psibar{\overline\psi}

\def\Dslash{D\kern-0.15em\raise0.17ex\llap{/}\kern0.15em\relax}
\def\Dslashl{\mathop{\Dslash}\limits^\leftarrow}
\def\Dmul{\mathop{D_\mu}\limits^\leftarrow}
\def\partialmul{\mathop{\partial_\mu}\limits^\leftarrow}
\def\Rl{\mathop{R}\limits^\leftarrow}

\def\Xl{\mathop{X}\limits^\leftarrow}
\def\Dl{\mathop{D}\limits^\leftarrow}
\def\sslash{s\kern-0.026em\raise0.17ex\llap{/}%
          \kern0.026em\relax}
\def\kslash{k\kern-0.026em\raise0.17ex\llap{/}%
          \kern0.026em\relax}
\def\pslash{p\kern-0.026em\raise0.17ex\llap{/}%
          \kern0.026em\relax}
\def\qslash{q\kern-0.026em\raise0.17ex\llap{/}%
          \kern0.026em\relax}
\def\tr{\mathop{\rm tr}}
%
\REF\HAS{%
P. Hasenfratz, V. Laliena and F. Niedermayer,
Phys.\ Lett.\ {\bf B427} (1998), 125.}
\REF\LUS{%
M. L\"uscher,
Phys.\ Lett.\ {\bf B428} (1998), 342.}
\REF\GIN{%
P. H. Ginsparg and K. G. Wilson,
Phys.\ Rev.\ {\bf D25} (1982), 2649.}
\REF\NEU{%
H. Neuberger,
Phys.\ Lett.\ {\bf B417} (1998), 141.}
\REF\NAR{%
R. Narayanan and H. Neuberger,
Nucl.\ Phys.\ {\bf B412} (1994), 574.}
\REF\KIK{%
Y. Kikukawa and A. Yamada,
Phys.\ Lett.\ {\bf B448} (1999), 265.}
\REF\FUJ{%
K. Fujikawa,
Nucl.\ Phys.\ {\bf B546} (1999), 480.}
\REF\HAG{%
K. Haga, H. Igarashi, K. Okuyama and H. Suzuki,
Phys.\ Rev.\ {\bf D55} (1997), 5325.}
\REF\SEI{%
E. Seiler and I. Stamatescu,
Phys.\ Rev.\ {\bf D25} (1982), 2177.}
\REF\AOK{%
S. Aoki and I. Ichinose,
Nucl.\ Phys.\ {\bf B272} (1986), 281.\nextline
S. Aoki,
Phys.\ Rev.\ {\bf D35} (1987), 1435.}
\REF\KAR{%
L. H. Karsten and J. Smit,
Nucl.\ Phys.\ {\bf B183} (1981), 103.}
\REF\RAN{%
S. Randjbar-Daemi and J. Strathdee,
Phys.\ Lett.\ {\bf B402} (1997), 134.}
\REF\CHI{%
T.-W. Chiu,
hep-lat/9810002.}
\REF\ADA{%
D. H. Adams,
hep-lat/9812003.}
%
\Pubnum={IU-MSTP/31; hep-th/9812019}
\date={November 1998}
\titlepage
\title{Simple Evaluation of the Chiral Jacobian with the Overlap Dirac
Operator}
\author{%
Hiroshi Suzuki\foot{E-mail: hsuzuki@mito.ipc.ibaraki.ac.jp}}
\address{%
Department of Physics, Ibaraki University, Mito 310-0056, Japan}
\abstract{%
The chiral Jacobian, which is defined with Neuberger's overlap Dirac
operator of the lattice fermion, is explicitly evaluated in the
continuum limit without expanding it in the gauge coupling constant.
Our calculational scheme is simple and straightforward. We determine
a coefficient of the chiral anomaly for general values of the mass
parameter and the Wilson parameter of the overlap Dirac operator.
}
\endpage
In a recent studies of the chiral symmetries on a lattice, the lattice
chiral Jacobian~[\HAS,\LUS]
$$
   \ln J=-2ia^4\sum_x\alpha(x)
   \tr\gamma_5\biggl[1-{1\over2}aD(x)\biggr]
   \delta(x,y)\biggr|_{y=x},
   \qquad\delta(x,y)={1\over a^4}\delta_{x,y},
\eqn\one
$$
associated with the Dirac operator~$D(x)$ which satisfies the
Ginsparg-Wilson relation~[\GIN]
$$
   D(x)\gamma_5+\gamma_5D(x)=aD(x)\gamma_5D(x),
\eqn\seven
$$
plays a central role. The simplest example of such a $D(x)$~is given
by the overlap Dirac operator~[\NEU,\NAR],
$$
   S=a^4\sum_x\psibar(x)iD(x)\psi(x),\qquad
   aD(x)=1+X(x){1\over\sqrt{X^\dagger(x)X(x)}},
\eqn\two
$$
where $X(x)$~is the Wilson-Dirac operator
$$
   X(x)=i\Dslash(x)-{m\over a}+R(x),
\eqn\three
$$
and the lattice covariant derivative~$D_\mu(x)$ and the Wilson
term~$R(x)$~are defined by
($\Dslash(x)=\sum_\mu\gamma^\mu D_\mu(x)$, and our gamma matrices are
anti-Hermitian: $\gamma^{\mu\dagger}=-\gamma^\mu$.)
$$
\eqalign{
   &D_\mu(x)={1\over2a}
   \Bigl[U_\mu(x)e^{a\partial_\mu}
   -e^{-a\partial_\mu}U_\mu^\dagger(x)\Bigr],
\cr
   &R(x)={r\over2a}
   \sum_\mu\Bigl[2-U_\mu(x)e^{a\partial_\mu}
   -e^{-a\partial_\mu}U_\mu^\dagger(x)\Bigr].
\cr
}
\eqn\four
$$
Note that $\Dslash^\dagger(x)=\Dslash(x)$ and $R^\dagger(x)=R(x)$, and
thus $X^\dagger(x)=-i\Dslash(x)-m/a+R(x)=\gamma_5X(x)\gamma_5$.

The chiral Jacobian~\one\ with $\alpha(x)=i/2$ represents the index
theorem on the lattice with finite lattice spacing~[\HAS] and it has
been shown to be an integer-valued topological invariant. It can also
be characterized as the Jacobian factor associated with the (local)
chiral $U(1)$ transformation~[\LUS],\foot{%
We have introduced the notation
$$
\eqalign{
   &a\Dl(x)=-1+\Xl(x){1\over\sqrt{{\Xl}^\dagger(x)\Xl(x)}},
\cr
   &\Xl(x)=i\Dslashl+{m\over a}-\Rl(x),\qquad
   {\Xl}^\dagger(x)=-i\Dslashl+{m\over a}-\Rl(x),
\cr
   &\Dmul(x)=-{1\over2a}
   \Bigl[U_\mu(x)e^{-a\partialmul}
   -e^{a\partialmul}U_\mu^\dagger(x)\Bigr],
\cr
   &\Rl(x)={r\over2a}
   \sum_\mu\Bigl[2-U_\mu(x)e^{-a\partialmul}
   -e^{a\partialmul}U_\mu^\dagger(x)\Bigr].
\cr
}
$$
}
$$
\eqalign{
   &\psi(x)\to
   \biggl\{1+i\alpha(x)\gamma_5\Bigl[1-{1\over2}aD(x)\Bigr]\biggr\}
   \psi(x),
\cr
   &\psibar(x)\to\psibar(x)
   \biggl\{1+i\Bigl[1+{1\over2}a\Dl(x)\Bigr]\gamma_5\alpha(x)\biggr\}.
\cr
}
\eqn\five
$$
Therefore, from the variation of the action~\two\ under~Eq.~\five,
$$
\eqalign{
   \delta S&=a^4\sum_x\alpha(x)
   \psibar(x)\Bigl[D(x)+\Dl(x)\Bigr]\gamma_5
   \Bigl[1-aD(x)\Bigr]\psi(x)
\cr
   &\equiv\int d^4x\,\alpha(x)\partial_\mu j_5^\mu(x),
\cr
}
\eqn\six
$$
(where use of the Ginsparg-Wilson relation~\seven\ has been made)
one has the Ward-Takahashi identity for the chiral anomaly:
$$
   \partial_\mu\VEV{j_5^\mu(x)}
   =2i\tr\gamma_5\biggl[1-{1\over2}aD(x)\biggr]
   \delta(x,y)\biggr|_{y=x}.
\eqn\eight
$$

The chiral $U(1)$ anomaly~\eight\ with the overlap Dirac
operator~\two\ has explicitly been calculated~[\KIK] in the continuum
limit by expanding the right-hand side with respect to the gauge
coupling constant (see also Ref.~[\GIN]). The authors of Ref.~[\KIK]
also observed that the coefficient of the chiral anomaly can be
expressed as a topological object in five-dimensional momentum space,
and thus it is stable with respect to a variation of the mass
parameter~$m$ and the Wilson parameter~$r$. When the overlap Dirac
operator~\two\ possesses only one massless pole~[\NEU], $0<m<2r$,
Eq.~\eight\ reproduces the correct magnitude in the continuum field
theory, $ig^2\varepsilon^{\mu\nu\rho\sigma}%
\tr F_{\mu\nu}F_{\rho\sigma}/16\pi^2$~[\KIK]. Also, quite recently,
it was shown~[\FUJ] that the continuum limit of the
expression~\eight\ generally displays the correct chiral anomaly
(including the coefficient), under general assumptions regarding the
Dirac operator, i.e., the Ginsparg-Wilson relation and the absence of
doublers.

In this note, we present a short evaluation of the chiral
Jacobian~\one\ or~\eight\ with the overlap Dirac operator~\two\
in the continuum limit~$a\to0$,\foot{%
The gauge field is treated as a non-dynamical background, and the
gauge field configuration is assumed to have a smooth continuum limit
(the well-defined spatial derivative).} but without an expansion in
the gauge coupling constant~$g$. This is possible because the chiral
anomaly is infrared finite and one may simply expand the
expression~\eight\ in the lattice spacing~$a$. Therefore, the actual
calculation does not require a perturbative expansion of the lattice
Dirac propagator in~$g$, which often becomes cumbersome. Our
calculational scheme is basically the same as that of~Ref.~[\HAG]
(similar calculational schemes can be found in~Refs.~[\SEI]
and~[\AOK]). We also determine the coefficient of the chiral anomaly
for general values of the parameters $m$ and~$r$. Although we present
the calculation for the overlap Dirac operator, the scheme itself is
not sensitive to the specific choice of the Dirac operator and, once
an explicit form of the Dirac operator is given, it provides a quick
way to evaluate the chiral Jacobian in the continuum limit.

First, we write the integrand of the Jacobian~\one, or the opposite of
the chiral anomaly~\eight, as follows (where we use~$\tr\gamma_5=0$)
$$
\eqalign{
   &i\tr\gamma_5aD(x)\delta(x,y)\Bigr|_{y=x}
\cr
   &=i\tr\gamma_5\Bigl(i\Dslash-{m\over a}+R\Bigr)
\cr
   &\quad\times
   \biggl[-\sum_\mu D_\mu^2
   +{1\over4}\sum_{\mu,\nu}[\gamma^\mu,\gamma^\nu][D_\mu,D_\nu]
   -i[\Dslash,R\,]+\Bigl({m\over a}-R\Bigr)^2\biggr]^{-1/2}
   \delta(x,y)\Bigr|_{y=x}.
\cr
}
\eqn\nine
$$
We next use $\delta(x,y)=%
\int_{-\pi}^{\pi}d^4k\,e^{ik(x-y)/a}/(2\pi a)^4$ and the relation
$$
\eqalign{
   &e^{-ikx/a}D_\mu e^{ikx/a}={i\over a}s_\mu+\widetilde D_\mu,
\cr
   &e^{-ikx/a}Re^{ikx/a}={r\over a}\sum_\mu(1-c_\mu)
   +\widetilde R,
\cr
}
\eqn\ten
$$
where $s_\mu=\sin k_\mu$ and $c_\mu=\cos k_\mu$ and
$$
\eqalign{
   &\widetilde D_\mu=
   {1\over2a}\Bigl[e^{ik_\mu}(U_\mu e^{a\partial_\mu}-1)
    -e^{-ik_\mu}(e^{-a\partial_\mu}U_\mu^\dagger-1)\Bigr],
\cr
   &\widetilde R=
   -{r\over2a}\sum_\mu\Bigl[e^{ik_\mu}(U_\mu e^{a\partial_\mu}-1)
    +e^{-ik_\mu}(e^{-a\partial_\mu}U_\mu-1)\Bigr].
\cr
}
\eqn\eleven
$$
Equation~\nine\ can then be written as
$$
\eqalign{
   &i\tr\gamma_5aD(x)\delta(x,y)\Bigr|_{y=x}
\cr
   &=-{i\over a^4}\int_{-\pi}^\pi
   {d^4k\over(2\pi)^4}\,\tr\gamma_5
   \Bigl[\sslash+m+r\sum_\mu(c_\mu-1)-ia\widetilde\Dslash
   -a\widetilde R\Bigr]
\cr
   &\qquad\times
   \biggl\{\sum_\nu(s_\nu-ia\widetilde D_\nu)^2
   +\Bigl[m+r\sum_\nu(c_\nu-1)-a\widetilde R\Bigr]^2
\cr
   &\qquad\qquad\qquad\qquad\qquad\qquad\qquad
   +{a^2\over2}\sum_{\nu,\rho}\gamma^\nu\gamma^\rho
   [\widetilde D_\nu,\widetilde D_\rho]
   -ia^2[\widetilde\Dslash,\widetilde R\,]\biggr\}^{-1/2}.
\cr
}
\eqn\twelve
$$
It is easy to find the $a\to0$ limit of this expression, because
$\gamma_5$ requires at least four gamma matrices
($\tr\gamma_5\gamma^\mu\gamma^\nu\gamma^\rho\gamma^\sigma%
=-4\varepsilon^{\mu\nu\rho\sigma}$ and $\varepsilon^{1234}=1$).
Finally, by parameterizing the link variable as
$U_\mu(x)=\exp[iagA_\mu(x)]$ and noting the relations
$$
   \widetilde D_\mu=c_\mu D_\mu^c+O(a),\qquad
   \widetilde R=-ir\sum_\mu s_\mu D_\mu^c+O(a),
\eqn\thirteen
$$
where $D_\mu^c=\partial_\mu+igA_\mu$ is the covariant derivative of
the continuum theory, and
$$
   [\widetilde D_\mu,\widetilde D_\nu]
   =igc_\mu c_\nu F_{\mu\nu}+O(a),\qquad
   [\widetilde D_\mu,\widetilde R\,]
   =grc_\mu\sum_\nu s_\nu F_{\mu\nu}+O(a),
\eqn\fourteen
$$
we find
$$
   i\lim_{a\to0}\tr\gamma_5aD(x)\delta(x,y)\Bigr|_{y=x}
   =-{ig^2\over16\pi^2}I(m,r)\varepsilon^{\mu\nu\rho\sigma}
   \tr F_{\mu\nu}F_{\rho\sigma}(x).
\eqn\fifteen
$$
The lattice integral~$I(m,r)$ is given by
$$
\eqalign{
   &I(m,r)
\cr
   &={3\over8\pi^2}\int_{-\pi}^\pi d^4k\,\prod_\mu c_\mu
   \biggl[m+r\sum_\nu(c_\nu-1)+r\sum_\nu{s_\nu^2\over c_\nu}\,\biggr]
\cr
   &\qquad\qquad\qquad\qquad\qquad\qquad\quad
   \times\biggl\{\sum_\rho s_\rho^2
   +\Bigl[m+r\sum_\rho(c_\rho-1)\Bigr]^2\biggr\}^{-5/2}
\cr
   &={3\over8\pi^2}\sum_{\epsilon_\mu=\pm1}
   \biggl(\prod_\mu\epsilon_\mu\biggr)\int_{-1}^1 d^4s
\cr
   &\qquad\times\Bigl\{
   m+r\sum_\nu\bigl[\epsilon_\nu(1-s_\nu^2)^{1/2}-1\bigr]
   +r\sum_\nu s_\nu^2\epsilon_\nu(1-s_\nu^2)^{-1/2}\Bigr\}
   B(s)^{-5/2},
\cr
}
\eqn\sixteen
$$
where
$$
   B(s)\equiv\sum_\mu s_\mu^2+
   \Bigl\{m+r\sum_\mu\bigl[\epsilon_\mu(1-s_\mu^2)^{1/2}-1\bigr]
   \Bigr\}^2.
\eqn\seventeen
$$
In writing the last expression of~Eq.~\sixteen, we have changed the
integration variable from $k_\mu$ to~$\sin k_\mu$ by splitting the
integration region into $-\pi/2\leq k_\mu<\pi/2$ and
$\pi/2\leq k_\mu<3\pi/2$ in each direction. Thus the original
integration region has been split into $2^4=16$~blocks; the four
vector~$\epsilon_\mu=(\pm1,\pm1,\pm1,\pm1)$ specifies the individual
block. Note that $\cos k_\mu$ is expressed as
$c_\mu=\epsilon_\mu(1-s_\mu^2)^{1/2}$.

Our next task is to evaluate the lattice integral~$I(m,r)$,
Eq.~\sixteen. To reproduce the correct coefficient of the chiral
anomaly for a single fermion from Eq.~\eight\ and~Eq.~\fifteen, one
would expect $I(m,r)=1$ at least in some parameter region. In fact,
the integral~$I(m,r)$ does not vary under an infinitesimal variation
of the parameters $m$ and~$r$. To see this, we note the identity
$$
\eqalign{
   &\Bigl\{
   m+r\sum_\mu\bigl[\epsilon_\mu(1-s_\mu^2)^{1/2}-1\bigr]
   +r\sum_\mu s_\mu^2\epsilon_\mu(1-s_\mu^2)^{-1/2}\Bigr\}
\cr
   &\qquad\qquad\qquad\qquad\qquad\qquad\quad
   \times\Bigl\{
   m+r\sum_\nu\bigl[\epsilon_\nu(1-s_\nu^2)^{1/2}-1\bigr]\Bigr\}
\cr
   &=B(s)
   +{1\over5}B(s)^{7/2}\sum_\mu s_\mu{\partial\over\partial s_\mu}
   B(s)^{-5/2},
\cr
}
\eqn\eighteen
$$
which is analogous to the identity utilized for the chiral anomaly of
the Wilson fermion~[\KAR,\SEI]. Having obtained Eq.~\eighteen, it is
straightforward to see
$$
   {\partial\over\partial m}I(m,r)
   =-{3\over8\pi^2}\sum_{\epsilon_\mu=\pm1}
   \biggl(\prod_\mu\epsilon_\mu\biggr)\int_{-1}^1 d^4s\,
   \biggl(4+\sum_\nu s_\nu{\partial\over\partial s_\nu}\biggr)
   B(s)^{-5/2},
\eqn\nineteen
$$
and, similarly,
$$
\eqalign{
   &{\partial\over\partial r}I(m,r)
\cr
   &=-{3\over8\pi^2}\sum_{\epsilon_\mu=\pm1}
   \biggl(\prod_\mu\epsilon_\mu\biggr)\int_{-1}^1 d^4s\,
   \biggl(4+\sum_\nu s_\nu{\partial\over\partial s_\nu}\biggr)
   \bigl[\epsilon_\rho(1-s_\rho^2)^{1/2}-1\bigr]B(s)^{-5/2}.
\cr
}
\eqn\twenty
$$
When $m\neq0$, $2r$, $4r$, $6r$ or~$8r$, $B(s)^{-5/2}$
in~Eqs.~\nineteen\ and~\twenty\ is regular in the integration region.
Therefore we can safely perform the partial integration, and
$\partial I(m,r)/\partial m%
=\partial I(m,r)/\partial r=0$ immediately follows. Note that the
surface terms are canceled out, because they are independent
of~$\epsilon_\mu$. This stability of the coefficient of the chiral
anomaly is consistent with the result of~Ref.~[\KIK] that the
coefficient of the triangle diagram for the chiral anomaly is
expressed as a topological object in the momentum space.

The above proof of the stability of~$I(m,r)$ indicates how we
should proceed: $I(m,r)$ can be regarded as a function of the ratio
of two parameters~$\alpha=m/r$ and the Wilson parameter~$r$. For fixed
$\alpha\neq0$, $2$, $4$, $6$ or~$8$, $I(m,r)$ is independent of the
value of~$r$. Therefore, we may evaluate it with a certain limiting
value of~$r$. We consider the limit~$r\to0$. By rescaling the
integration variable in~Eq.~\sixteen\ as $s_\mu\to rs_\mu$, we have 
$$
\eqalign{
   I(\alpha r,r)&={3\over8\pi^2}\sum_{\epsilon_\mu=\pm1}
   \biggl(\prod_\mu\epsilon_\mu\biggr)
   \int_{-1/r}^{1/r}d^4s
\cr
   &\qquad\qquad\quad\times
   {\alpha+\sum_\nu\bigl[\epsilon_\nu(1-r^2s_\nu^2)^{-1/2}-1\bigr]
   \over\Bigl(\sum_\rho s_\rho^2
   +\Bigl\{\alpha
   +\sum_\rho\bigl[\epsilon_\rho(1-r^2s_\rho^2)^{1/2}-1\bigr]
   \Bigr\}^2\Bigr)^{5/2}}.
\cr
}
\eqn\twentyone
$$
To consider the $r\to0$~limit, we divide the integration region
$[-1/r,1/r]^4$ of~Eq.~\twentyone\ into a four-dimensional
cylinder~$C(L)\equiv S^3\times[-L,L]$ and the
remaining~$R(L)\equiv[-1/r,1/r]^4-C(L)$. The radius of~$S^3$ is~$L$
($L<1/r$), and the direction of the cylinder is taken along the
$\nu$-direction in the numerator. Then it is possible to show that
the $r\to0$~limit of the latter integral, $I(\alpha r,r)_{R(L)}$,
vanishes as~$L\to\infty$.\foot{%
The proof goes as follows: The absolute value of the integral
over~$R(L)$, $\bigl|I(\alpha r,r)_{R(L)}\bigr|$, can be bounded
by a linear combination of
$$
   4\pi\int_0^{\sqrt{3}/r}d\rho\,\rho^2\int_L^{1/r}dz\,
   \left\{{1\atop(1-r^2z^2)^{-1/2}}\right\}(\rho^2+z^2)^{-5/2}
$$
and $\int_L^{\sqrt{3}/r}d\rho\,\int_{-L}^Ldz$ with the same integrand.
After integrating over~$\rho$, this integral is bounded by 
$$
\eqalign{
   &{4\pi\over3}\int_L^{1/r}dz\,
   \left\{{z^{-2}\atop z^{-2}(1-r^2z^2)^{-1/2}}\right\}
   (1+r^2z^2/3)^{-3/2}
   <{4\pi\over3}\int_L^{1/r}dz\,
   \left\{{z^{-2}\atop z^{-2}(1-r^2z^2)^{-1/2}}\right\}
\cr
   &={4\pi\over3}\left\{-r+L^{-1}\atop(1-r^2L^2)^{1/2}L^{-1}\right\}
   \buildrel{r\to0}\over\longrightarrow{4\pi\over3L}.
\cr
}
$$
Similarly, the integral over
$\int_L^{\sqrt{3}/r}d\rho\,\int_{-L}^Ldz$ has a bound whose limit as
$r\to0$ is~$4\pi/L$. Therefore, we have
$$
   \bigl|\lim_{r\to0}I(\alpha r,r)_{R(L)}\bigr|
   <{{\rm const}\over L}.
$$
} Therefore, the $r\to0$~limit of the integral~\twentyone\ can be
evaluated by restricting the integration region to~$C(L)$, by taking
the $L\to\infty$~limit. Namely,
$$
\eqalign{
   \lim_{r\to0}I(\alpha r,r)
   &={3\over8\pi^2}\sum_{\epsilon_\mu=\pm1}
   \biggl(\prod_\mu\epsilon_\mu\biggr)\lim_{L\to\infty}
   \lim_{r\to0}\int_{-L}^{L}d^4s
\cr
   &\qquad\qquad\quad\times
   {\alpha+\sum_\nu\bigl[\epsilon_\nu(1-r^2s_\nu^2)^{-1/2}-1\bigr]
   \over\Bigl(\sum_\rho s_\rho^2
   +\Bigl\{\alpha
   +\sum_\rho\bigl[\epsilon_\rho(1-r^2s_\rho^2)^{1/2}-1\bigr]
   \Bigr\}^2\Bigr)^{5/2}}
\cr
   &={3\over8\pi^2}\sum_{\epsilon_\mu=\pm1}
   \biggl(\prod_\mu\epsilon_\mu\biggr)\int_{-\infty}^\infty d^4s\,
   {\alpha+\sum_\nu(\epsilon_\nu-1)
   \over\Bigl\{\sum_\rho s_\rho^2
   +\bigl[\alpha+\sum_\rho(\epsilon_\rho-1)\bigr]^2\Bigr\}^{5/2}}
\cr
   &={1\over2}\sum_{\epsilon_\mu=\pm1}
   \biggl(\prod_\mu\epsilon_\mu\biggr)
   {\alpha+\sum_\nu(\epsilon_\nu-1)
   \over|\alpha+\sum_\rho(\epsilon_\rho-1)|}.
}
\eqn\add
$$
In this way, the lattice integral~\sixteen\ is given by
$$
\eqalign{
   I(m,r)&=\lim_{r\to0}I(\alpha r,r)
\cr
   &=\theta(m/r)-4\theta(m/r-2)+6\theta(m/r-4)
   -4\theta(m/r-6)+\theta(m/r-8),
\cr
}
\eqn\twentytwo
$$
where $\theta(x)$ is the step function. As anticipated, the
integral~$I(m,r)$ varies only stepwise depending on the ratio~$m/r$.
Finally, by recalling Eqs.~\fifteen\ and~\one, we have the chiral
Jacobian in the continuum limit,
$$
\eqalign{
   &\lim_{a\to0}\ln J
\cr
   &=-{ig^2\over16\pi^2}
   \bigl[\theta(m/r)-4\theta(m/r-2)+6\theta(m/r-4)
   -4\theta(m/r-6)+\theta(m/r-8)\bigr]
\cr
   &\qquad\qquad\qquad\qquad\qquad\qquad\qquad\qquad\qquad
   \times\int d^4x\,\alpha(x)
   \varepsilon^{\mu\nu\rho\sigma}\tr F_{\mu\nu}F_{\rho\sigma}(x).
\cr
}
\eqn\twentythree
$$
It is interesting to note that the quantity inside the square brackets
has a simple physical meaning: It is sum of the chiral charge~[\KAR]
of the massless degrees of freedom. In particular, when the Dirac
operator has only one massless pole~[\NEU], $0<m<2r$,
Eq.~\twentythree\ reproduces the correct coefficient as a single
fermion, as was first shown in~Ref.~[\KIK]. For other ranges of~$m$,
our expression~\twentythree\ completely agrees with the general
result of the analysis of~Ref.~[\FUJ] of the overlap Dirac operator.

The author is grateful to Professor K.~Fujikawa for helpful
suggestions and also to reviewers for useful comments. This work is
supported in part by the Ministry of Education Grant-in-Aid for
Scientific Research, Nos.~09740187 and~10120201.

\bigskip
\titlestyle{\bf Note added:}
The detailed analysis of chiral (gauge) anomalies in the overlap
formulation in general is performed in~Ref.~[\RAN]. The dynamical
implication of the special values of the mass parameter $m$,
$m=0$, $2r$, $4r$, $6r$ and~$8r$ is fully investigated
in~Ref.~[\CHI]. I would like to thank the authors of these papers for
information. Almost the same result as ours is obtained independently
by Dr.~D.~H.~Adams in~Ref.~[\ADA].

\refout

\bye